\documentstyle[aps,prl,epsf,amssymb,amsmath,amsthm]{revtex}

\bibstyle{unsrt}

\newcommand{\half}{\mbox{$\textstyle \frac{1}{2}$}}

\newcommand{\octa}{\mbox{$\textstyle \frac{1}{8}$}}
\newcommand{\rd}{{\rm d}} \newcommand{\ri}{{\rm i}}
\newcommand{\re}{{\rm e}}

\tighten
\begin{document}
\draft \preprint{}

\twocolumn[\hsize\textwidth\columnwidth\hsize\csname
@twocolumnfalse\endcsname

\title{Finite-time stochastic reduction models}

\author{Dorje C. Brody$^*$ and Lane P. Hughston$^\dagger$}

\address{${}^*$Blackett Laboratory, Imperial College, London SW7
2BZ, UK}
\address{${}^\dagger$Department of
Mathematics, King's College London, The Strand, London WC2R 2LS,
UK}

\date{\today}
\maketitle

\begin{abstract}
A new energy-based stochastic extension of the Schr\"odinger
equation for which the wave function collapses after the passage
of a finite amount of time is proposed. An exact closed-form
solution to the dynamical equation, valid for all
finite-dimensional quantum systems, is presented and used to
verify explicitly that reduction of the state vector to an energy
eigenstate occurs.  A time-change technique is introduced to
construct a `clock' variable that relates the asymptotic and the
finite-time collapse models by means of a nonlinear
transformation.
\end{abstract}

\vskip2pc]

The idea that Brownian motion might play a role in models for the
collapse of the wave function was first envisaged by Wiener and
Siegel~\cite{wiener}. Although that specific proposal did not in
the end fully account for the correct probability law, a number of
authors have subsequently proposed and developed dynamical
collapse models driven by Brownian motion that have the key
property of being compatible with the probabilistic hypotheses of
standard quantum mechanics. Progress in this area over the last
two decades can be broadly classified into two categories: (a)
models based on the idea of `spontaneous localisation' of the wave
function~\cite{grw}, and (b) models based on the notion of
collapse of the state vector to an energy
eigenstate~\cite{energy}. See~\cite{review} and references cited
therein for accounts of these approaches.

The energy-based collapse models, with which we are concerned
here, have the physical property that the expectation of the
future energy of the system is given by the initial energy. It has
been argued on phenomenological grounds \cite{hughston} that the
characteristic timescale for collapse to an energy eigenstate in
such models is of the order
\begin{eqnarray}
\tau_{R} \approx \left( \frac{2.8{\rm MeV}} {\Delta H}\right)^2
{\rm s}, \label{eq:1}
\end{eqnarray}
where $\Delta H$ is the initial energy uncertainty, and it has
been demonstrated that the choice (\ref{eq:1}) is consistent with
empirical observations for a number of different examples of
quantum systems~\cite{adler}. The timescale $\tau_{R}$ is
indicative of the time it takes for the wave function to reach the
immediate vicinity of an energy eigenstate. That is to say, after
the passage of several multiples of $\tau_R$, the state of the
system is, with a high degree of probability, nearly
indistinguishable from one of the energy eigenstates. It should be
emphasised, however, that strict collapse, in these models, is
achieved only asymptotically in time, and it has hitherto been
unknown how to formulate a consistent dynamical collapse model
that exhibits a complete reduction of the state vector in a finite
period of time.

The purpose of this article is to introduce a new class of
energy-based models for which the collapse is completed after the
passage of a specified time interval $T$. Furthermore, we obtain
an exact closed-form expression for the state vector process that
solves the dynamical equation. We show that, remarkably, the
finite-time collapse model and the standard infinite-time collapse
models are related by a time transformation of the form
\begin{eqnarray}
\tau(t) = \frac{tT}{T-t}, \label{eq:2}
\end{eqnarray}
where $t$ is the clock time of the finite-time collapse model, and
$\tau$ is the clock time in the asymptotic collapse model. Thus
the finite-time collapse model can be viewed as a `fast-forwarded'
version of the asymptotic collapse model.

We begin by stating the main results of the paper. {\it Let the
dynamics of the state-vector process $\{|\psi_t\rangle\}$ be given
by the following stochastic Schr\"odinger equation:
\begin{eqnarray}
\rd |\psi_t\rangle &=& -\ri{\hat H}|\psi_t\rangle\rd t - \octa
\sigma_t^2 ({\hat H}-H_t )^2|\psi_t\rangle\rd t \nonumber \\ &&
+\half \sigma_t({\hat H}-H_t)|\psi_t\rangle\rd W_t. \label{eq:3}
\end{eqnarray}
Here $H_t = \langle{\psi}_t|{\hat H}|\psi_t\rangle/
\langle{\psi}_t|\psi_t\rangle$ is the expectation value of the
Hamiltonian ${\hat H}$ in the state $|\psi_t\rangle$, $\{W_t\}$ is
the Wiener process, and $\sigma_t =\sigma T/(T-t)$, where $\sigma$
is a parameter. Then starting from an arbitrary initial state
$|\psi_0\rangle$ the wave function collapses to an energy
eigenstate at time $T$. Energy is conserved in expectation, and
the probability of collapse to a state of energy $E_i$ {\rm
(}$i=1,2,3,\ldots,n${\rm )} is in accordance with the Born law
$\pi_i=\langle\psi_0|{\hat\Pi}_i|\psi_0\rangle/ \langle
\psi_0|\psi_0\rangle$, where ${\hat\Pi}_i$ is the projection
operator onto the Hilbert subspace of states with energy $E_i$.}

Our model (\ref{eq:3}) contains two parameters, namely, the
reduction time $T$ and the energy volatility parameter $\sigma$,
which has the units $[{\rm energy}]^{-1} [{\rm time}]^{-1/2}$. An
application of the Ito product rule
\begin{eqnarray}
\rd(X_tY_t)=X_t(\rd Y_t)+(\rd X_t)Y_t+(\rd X_t)(\rd Y_t)
\label{eq:33}
\end{eqnarray}
shows that the normalisation of $|\psi_t\rangle$ is preserved
under (\ref{eq:3}), so $\rd\langle\psi_t|\psi_t\rangle=0$, and
that $\{H_t\}$ satisfies
\begin{eqnarray}
\rd H_t = \sigma_t V_t\,\rd W_t, \label{eq:5}
\end{eqnarray}
where
\begin{eqnarray}
V_t=\frac{\langle{\psi}_t|({\hat H}-H_t)^2|\psi_t\rangle}
{\langle\psi_t|\psi_t\rangle}
\end{eqnarray}
is the energy variance. It follows from (\ref{eq:5}) that the
energy process $\{H_t\}$ has no drift and thus satisfies the
conservation law
\begin{eqnarray}
{\mathbb E}[ H_u|\{W_s\}_{0\leq s\leq t}]=H_t \label{eq:6}
\end{eqnarray}
for $0\leq t\leq u$, where ${\mathbb E}[-|\{W_s\}_{0\leq s\leq
t}]$ denotes conditional expectation given the trajectory of the
Wiener process from time $0$ up to time $t$.

To verify the collapse property we shall examine the dynamics of
the energy variance process $\{V_t\}$. In particular, by use of
the Ito rule (\ref{eq:33}) together with Jensen's inequality we
deduce that
\begin{eqnarray}
{\bar V}_t \leq V_0 - \int_0^t \sigma_s^2 {\bar V}_s^2 \rd s,
\label{eq:6.1}
\end{eqnarray}
where ${\bar V}_t={\mathbb E}[V_t]$ is the unconditional
expectation of the energy variance, which is non-negative. Since
${\bar V}_t\leq{\bar V}_s$ for $s\leq t$, the inequality
(\ref{eq:6.1}) remains valid if we replace ${\bar V}_s$ with
${\bar V}_T$ in the integrand. It follows for any $t\leq T$ that
\begin{eqnarray}
{\bar V}_t \leq V_0 - \sigma^2\frac{tT}{T-t} {\bar V}_T^2.
\label{eq:6.2}
\end{eqnarray}
Now suppose ${\bar V}_T>0$. Then there would exist a time $t_0=
V_0T/(V_0+\sigma^2T{\bar V}_T^2)<T$ such that the right side of
(\ref{eq:6.2}) vanishes and hence such that ${\bar V}_{t_0}= 0$.
This contradicts our supposition, and hence we conclude that
${\bar V}_T=0$, and thus $V_T=0$ with probability one.

To verify the Born law we must show that the conditional
transition probability
\begin{eqnarray}
\pi_{it}=\frac{\langle\psi_t|{\hat\Pi}_i|\psi_t\rangle}
{\langle\psi_t|\psi_t\rangle} \label{eq:44}
\end{eqnarray}
is conserved in expectation, i.e. that $\pi_{i}={\mathbb
E}[\pi_{it}]$ for $t\leq T$. This follows from an application of
Ito's rule, which implies that $\rd\pi_{it} =\sigma_t
(E_i-H_t)\pi_{it} \rd W_t$ and thus $\{\pi_{it}\}$ has no drift.
Because ${\mathbb E}[\pi_{iT}]$ is the probability of collapse to
a state with energy $E_i$, the conservation law shows that this is
given by the Born probability $\pi_i$.

We now proceed to investigate the dynamical equation (\ref{eq:3})
more deeply with a view to obtaining a closed-form solution. First
we observe that (\ref{eq:3}) can be cast into an integral form
that incorporates the initial condition. In particular, it follows
from (\ref{eq:3}) that
\begin{eqnarray}
|\psi_t\rangle = {\hat U}_t {\hat M}_t^{1/2}|\psi_0\rangle,
\label{eq:55}
\end{eqnarray}
where ${\hat U}_t={\rm e}^{-{\rm i} {\hat H}t}$ is the standard
unitary evolutionary operator associated with ${\hat H}$, and
\begin{eqnarray}
{\hat M}_t\!=\!\frac{\exp\left({\hat H}\int_0^t\!\sigma_s({\rd}W_s
+ \sigma_s H_s{\rd}s)\!-\!\frac{1}{2} {\hat H}^2 \int_0^t\!
\sigma_s^2 {\rd}s \right)}{\exp\left( \int_0^t\! \sigma_s H_s
({\rd} W_s+\sigma_s H_s{\rd}s)\!-\!\frac{1}{2} \int_0^t\!
\sigma_s^2 H_s^2 {\rd}s \right)} \label{eq:7}
\end{eqnarray}
is a positive operator-valued process. Next we introduce a process
$\{W_t^*\}$ defined by
\begin{eqnarray}
W_t^* = W_t + \int_0^t \sigma_s H_s {\rd}s. \label{eq:8}
\end{eqnarray}
With respect to the Wiener measure ${\mathbb P}$, under which
$\{W_t\}$ is a standard Brownian motion, $\{W_t^*\}$ is a Brownian
motion with drift. Therefore, by the well-known theorem of
Girsanov there exists a measure ${\mathbb Q}$ with the property
that $\{W_t^*\}$ is a ${\mathbb Q}$-Brownian motion for
$t\in[0,T)$. Since the dynamical law (\ref{eq:3}) preserves the
normalisation of $|\psi_t\rangle$, it follows from (\ref{eq:55})
that $\langle\psi_0|{\hat M}_t|\psi_0\rangle=\langle\psi_0|
\psi_0\rangle$ for $t\in[0,T)$, and hence that
\begin{eqnarray}
{\hat M}_t = \frac{1}{\Phi_t} \exp\left( {\hat H}\int_0^t\sigma_s
\rd W_s^* - \half {\hat H}^2\int_0^t\sigma_s^2 \rd s \right),
\label{eq:9}
\end{eqnarray}
where
\begin{eqnarray}
\Phi_t &=& \exp\left( \int_0^t \sigma_s H_s {\rd}W_s^* - \half
\int_0^t \sigma_s^2 H_s^2 {\rd}s \right) \nonumber \\ &=& \frac{
\langle \psi_0| \exp\left( {\hat H} \int_0^t\! \sigma_s \rd W_t^*
\!-\! \frac{1}{2} {\hat H}^2 \int_0^t\! \sigma_s^2 \rd t \right)
|\psi_0\rangle}{\langle\psi_0| \psi_0\rangle} . \label{eq:10}
\end{eqnarray}

We recall that at time $t$ the probability of reduction to a state
with energy $E_i$ is given by (\ref{eq:44}). By use of
(\ref{eq:55}) and the commutation relation $[{\hat
M}_t,{\hat\Pi}_i]=0$ it follows then that $\pi_{it}=\langle\psi_0|
{\hat\Pi}_i {\hat M}_t|\psi_0 \rangle/\langle\psi_0| \psi_0
\rangle$. Substituting (\ref{eq:9}) into this relation we thus
obtain
\begin{eqnarray}
\pi_{it} &=& \frac{\langle{\psi}_0| {\hat\Pi}_i \exp\left({\hat H}
\int_0^t\!\sigma_s\rd W_s^*\!-\!\frac{1}{2} {\hat H}^2\int_0^t\!
\sigma_s^2 \rd s\right)|\psi_0\rangle} {\langle\psi_0| \exp\left(
{\hat H}\int_0^t\!\sigma_s\rd W_s^*\!-\!\frac{1}{2} {\hat H}^2
\int_0^t\!\sigma_s^2 \rd s \right)|\psi_0\rangle} \nonumber \\ &=&
\frac{\pi_i \exp\left(E_i\int_0^t\sigma_s\rd W_s^* - \frac{1}{2}
E_i^2 \int_0^t\sigma_s^2 \rd s\right)}{\sum_i \pi_i \exp\left( E_i
\int_0^t\sigma_s\rd W_s^* -\frac{1}{2} E_i^2 \int_0^t \sigma_s^2
\rd s\right)}. \label{eq:15}
\end{eqnarray}
Recalling that $\sigma_s=\sigma T/(T-s)$, we find it convenient
now to introduce a process $\{\xi_t\}$ defined by
\begin{eqnarray}
\xi_t = (T-t)\int_0^t \frac{1}{T-s}\,\rd W_s^* . \label{eq:17}
\end{eqnarray}
We observe that the right side of (\ref{eq:17}) is an integral
representation for a \emph{Brownian bridge}~\cite{yor} in the
${\mathbb Q}$-measure. Substituting (\ref{eq:17}) into
(\ref{eq:15}) we infer that the reduction probability $\pi_{it}$
can be expressed in terms of the variable $\xi_t$ as follows:
\begin{eqnarray}
\pi_{it} = \frac{\pi_i \exp\left(\frac{\sigma\xi_t E_iT-
\frac{1}{2}\sigma^2 E_i^2 t T}{T-t}\right)}{\sum_i \pi_i
\exp\left(\frac{\sigma \xi_t E_iT- \frac{1}{2}\sigma^2 E_i^2 t
T}{T-t}\right)}. \label{eq:29.6}
\end{eqnarray}
Furthermore, on account of the relation $H_t=\sum_i E_i \pi_{it}$
for the energy process, we deduce that:
\begin{eqnarray}
H_t = \frac{\sum_i \pi_i E_i \exp\left(\frac{\sigma\xi_t E_iT-
\frac{1}{2}\sigma^2 E_i^2 t T}{T-t}\right)}{\sum_i \pi_i
\exp\left(\frac{\sigma \xi_t E_iT- \frac{1}{2}\sigma^2 E_i^2 t
T}{T-t}\right)} . \label{eq:29.5}
\end{eqnarray}

In view of the exact solution to the asymptotic collapse model
obtained in \cite{brody}, the fact that $H_t$ and $\pi_{it}$ can
be expressed as functions of $\xi_t$ and $t$ suggests there might
be a simple representation for $\xi_t$ in terms of elementary
random data that can be specified without reference to the
state-vector process. If so, we can substitute such a
representation for $\xi_t$ into (\ref{eq:29.5}) to obtain a
closed-form solution to the stochastic equation, in place of the
integral representation (\ref{eq:55}) which implicitly depends on
$\{|\psi_t \rangle\}$.

We thus proceed as follows: we define a process $\{\beta_t\}$ by
the relation $\beta_t=\xi_t-\sigma t H_T$, where $H_T$ is the
terminal value of the energy. We claim that the random variables
$\beta_t$ and $H_T$ are independent. To establish their
independence it suffices to verify that ${\mathbb E}[{\rm
e}^{x\beta_t+yH_T}] = {\mathbb E}[{\rm e}^{x\beta_t}]{\mathbb
E}[{\rm e}^{yH_T}]$ for arbitrary $x,y$. Using the tower property
of conditional expectation we have
\begin{eqnarray}
{\mathbb E}[{\rm e}^{x\beta_t+yH_T}] = {\mathbb E}\left[ {\rm
e}^{x\xi_t}\,{\mathbb E}\left[{\rm e}^{(y- \sigma tx) H_T}\Big|
\xi_t \right]\right] . \label{eq:19}
\end{eqnarray}
Let us consider the inner expectation ${\mathbb E} \left[ {\rm
e}^{(y- \sigma tx) H_T}|\xi_t\right]$. Using expression
(\ref{eq:29.6}) for the conditional probability distribution of
the terminal energy $H_T$ we obtain
\begin{eqnarray}
{\mathbb E} \left[ {\rm e}^{(y- \sigma tx) H_T}\Big| \xi_t \right]
&& \label{eq:20} \\ && \hspace{-2.7cm} = \Phi_t^{-1} \sum_i \pi_i
{\rm e}^{(y- \sigma tx)E_i} \exp\left( \frac{\sigma\xi_t E_iT-
\frac{1}{2} \sigma^2 E_i^2 t T}{T-t}\right). \nonumber
\end{eqnarray}
Now $\{\Phi_t\}$ is the density process for changing the measure
from ${\mathbb Q}$ to ${\mathbb P}$. That is to say, the
expectation ${\mathbb E}$ under ${\mathbb P}$, in which $\{W_t\}$
is a standard Brownian motion, is related to the expectation
${\mathbb E}^{\mathbb Q}$ under ${\mathbb Q}$, in which
$\{W_t^*\}$ is a standard Brownian motion, according to
\begin{eqnarray}
{\mathbb E}[X_u|\{W_s\}_{0\leq s\leq t}]= \frac{1}
{\Phi_t}{\mathbb E}^{\mathbb Q} [\Phi_u X_u|\{W_s\}_{0\leq s\leq
t}], \label{eq:21}
\end{eqnarray}
for any random variable $X_u$ that can be expressed as a
functional of the trajectory $\{W_s\}_{0\leq s\leq u}$. Then
making use of the fact that $\{\xi_t\}$ is a Brownian bridge under
${\mathbb Q}$, we deduce, after some rearrangement of terms, that
\begin{eqnarray}
{\mathbb E}\left[{\rm e}^{x\beta_t+yH_T}\right] = \sum_i \pi_i
{\rm e}^{yE_i}\, {\rm e}^{\frac{t(T-t)}{2T}x^2}. \label{eq:22}
\end{eqnarray}
Here we have used the facts that if $g$ is a zero-mean Gaussian
random variable with variance $\gamma^2$, then ${\mathbb E}[{\rm
e}^{xg}]={\rm e}^{\frac{1}{2}\gamma^2x^2}$, and that the variance
of the Brownian bridge $\{\xi_t\}$ is $t(T-t)/T$. This proves the
independence of $\{\beta_t\}$ and $H_T$.

The result (\ref{eq:22}) also establishes that under ${\mathbb P}$
the process $\{\beta_t\}$ is Gaussian, and has mean zero and
variance $t(T-t)/T$ for $t\in[0,T)$. A similar line of argument
shows for $s\leq t$ that the covariance of $\beta_s$ and $\beta_t$
is
\begin{eqnarray}
{\rm Cov}\left[\beta_s,\beta_t\right] = s(T-t)/T.
\end{eqnarray}
Therefore we conclude that $\{\beta_t\}$ is a ${\mathbb
P}$-Brownian bridge. As a consequence, there exists a Brownian
motion $\{B_t\}$ such that $\{\beta_t\}$ admits an integral
representation:
\begin{eqnarray}
\beta_t = (T-t) \int_0^t \frac{1}{T-s}\,\rd B_s . \label{eq:24}
\end{eqnarray}
Thus we have verified that the process $\{\xi_t\}$ can be
expressed in the form
\begin{eqnarray}
\xi_t = \sigma t H_T + \beta_t, \label{eq:25}
\end{eqnarray}
where $H_T$ is the terminal energy, and $\{\beta_t\}$ is an
independent Brownian bridge.

Conversely, {\it given a discrete random variable $H_T$ taking the
values $\{E_i\}$ with probability $\{\pi_i\}$, and given an
independent Brownian bridge process $\{\beta_t\}$ on the interval
$[0,T]$, we can use the ansatz {\rm (\ref{eq:25})} as a basis for
constructing a closed-form solution to the dynamical equation {\rm
(\ref{eq:3})}}.

The argument can be sketched as follows. First, given $\{\xi_t\}$
as defined above in terms of $H_T$ and $\{\beta_t\}$, let
$\pi_{it}={\mathbb P}(H_T=E_i|\xi_t)$ be the conditional
probability that $H_T$ takes the value $E_i$ when the value of
$\xi_t$ is specified. It follows by use of the Bayes formula that
\begin{eqnarray}
{\mathbb P}(H_T=E_i|\xi_t) = \frac{\pi_i \rho(\xi_t|H_T=E_i)}
{\sum_i \pi_i \rho(\xi_t|H_T=E_i)}, \label{eq:27}
\end{eqnarray}
where
\begin{eqnarray}
\rho(\xi_t|H_T=E_i)=\textstyle{\sqrt{\frac{T}{2\pi t(T-t)}}}
\exp\left( - \frac{(\xi_t-\sigma t E_i)^2T}{2t(T-t)}\right).
\label{eq:28}
\end{eqnarray}
Expression (\ref{eq:28}) can be deduced from the fact that
conditional on $H_T=E_i$ the variable $\xi_t$ in (\ref{eq:25}) is
normally distributed with mean $\sigma t E_i$ and variance
$t(T-t)/T$. Thus, we recover the expression (\ref{eq:29.6}) for
$\{\pi_{it}\}$. Then by an application of Ito calculus we obtain
\begin{eqnarray}
\rd  \pi_{it} = \sigma_t (E_i-H_t)\pi_{it}\,\rd W_t ,
\label{eq:30}
\end{eqnarray}
where now the process $\{W_t\}$ is {\it defined} in terms of
$\{\xi_t\}$ by the relation
\begin{eqnarray}
W_t = \int_0^t \frac{1}{T-s}\Big(\xi_s-\sigma TH_s\Big)\rd s +
\xi_t, \label{eq:31}
\end{eqnarray}
and $\{H_t\}$ is defined by $H_t=\sum_i\pi_{it}E_i$. The process
$\{W_t\}$ thus determined turns out to be a standard Brownian
motion. This can be demonstrated by showing that $\{W_t\}$
satisfies ${\mathbb E}[W_t|\{W_u\}_{0\leq u\leq s}]=W_s$ for
$s\leq t$ and $(\rd W_t)^2=\rd t$. Finally, if we write
\begin{eqnarray}
|\phi_i\rangle = \frac{{\hat\Pi}_i|\psi_0\rangle}{\langle\psi_0|
{\hat\Pi}_i| \psi_0\rangle^{1/2}}
\end{eqnarray}
for the standard L\"uders states, then the state vector
$\{|\psi_t\rangle\}$ that solves (\ref{eq:3}) can be written in
the form
\begin{eqnarray}
|\psi_t\rangle = {\rm e}^{-\ri{\hat H}t} \sum_i \sqrt{\pi_{it}}\,
|\phi_i\rangle,
\end{eqnarray}
or, more explicitly,
\begin{eqnarray}
|\psi_t\rangle = \sum_i \frac{\sqrt{\pi_{i}} \re^{-\ri E_i t +
\frac{1}{2} E_i \sigma_t \xi_t - \frac{1}{4} E_i^2 \int_0^t
\sigma_s^2 \rd s }}{\sqrt{ \sum_{i} \pi_{i} \re^{\frac{1}{2} E_i
\sigma_t \xi_t - \frac{1}{4} E_i^2 \int_0^t \sigma_s^2 \rd s
}}}|\phi_i\rangle, \label{eq:35}
\end{eqnarray}
where $\sigma_s=\sigma T/(T-s)$, and $\xi_t=\sigma t H_T +
\beta_t$. A straightforward calculation shows that $|\psi_t\rangle
\to |\phi_k\rangle$ as $t\to T$, if we set $H_T=E_k$ in
(\ref{eq:35}). By taking the stochastic differential of
(\ref{eq:35}), and using (\ref{eq:31}), we are led back to the
starting point (\ref{eq:3}), which directly verifies that
(\ref{eq:35}) is the solution to the dynamical equation
(\ref{eq:3}).

We thus conclude that the substitution of (\ref{eq:25}) in
(\ref{eq:29.5}) yields a closed-form expression for the energy
process in terms of the exogenously specifiable independent data
$\{H_T\}$ and $\{\beta_t\}$. As a consequence, we are able to
directly simulate the solution to the stochastic equation, as well
as the associated energy evolution $\{H_t\}$, for an arbitrary
finite-dimensional quantum system.

Next, we remark on the interpretation of the finite-time collapse
model (\ref{eq:3}). First we observe (cf. \cite{yor}) that if
$\{B_t\}$ is a standard Brownian motion on the interval $[0,T]$,
then the process $\{{\tilde B}_\tau\}$ defined by
\begin{eqnarray}
{\tilde B}_{\tau} = \int_0^{\frac{\tau T}{\tau+T}} \frac{T}{T-s}
\,\rd B_s \label{eq:36}
\end{eqnarray}
is a standard Brownian motion on the time interval $[0,\infty)$.
In particular, let us introduce a new time variable by the
relation $\tau(t)=tT/(T-t)$, and define
\begin{eqnarray}
\eta_\tau = \frac{T}{T-t}\,\xi_t = \left(1+\frac{\tau}{T}\right)
\, \xi_{\frac{\tau T}{\tau +T}}. \label{eq:37}
\end{eqnarray}
Then the equation (\ref{eq:25}) can be put into the form
\begin{eqnarray}
\eta_\tau = \frac{\sigma t T H_T}{T-t} + \frac{T\beta_t}{T-t} =
\sigma \tau H_T + \int_0^t \frac{T}{T-s}\,\rd B_s, \label{eq:38}
\end{eqnarray}
on account of the integral representation (\ref{eq:24}). Making
use of the relations (\ref{eq:36}) and $t=\tau T/(\tau+T)$, we
thus deduce that
\begin{eqnarray}
\eta_\tau = \sigma \tau H_T + {\tilde B}_\tau.
\end{eqnarray}
However, this is the ansatz (cf. \cite{brody}) that determines
{\it asymptotic} collapse to an energy eigenstate. That is to say,
the expectation $H_\tau={\mathbb E}[H_T| \eta_\tau]$ determines
the energy process $\{H_\tau\}$ associated with the standard
energy-based infinite-time collapse models. Therefore, if we take
the infinite-time collapse model and replace the time variable $t$
with a clock $\tau(t)$, then as $t\to T$, the clock $\tau(t)$
`ticks' faster and faster. Eventually, in a system for which time
is measured by `internal' time $t$, the collapse takes place in a
finite time interval $T$, whereas in a system for which time is
measured by the clock variable $\tau(t)$, the collapse takes place
asymptotically as $\tau \to\infty$.

In summary, we have introduced a consistent energy-based collapse
model that achieves state reduction in finite time, thus meeting
the `challenge' posed in Ref.~\cite{gisin}. We have verified the
collapse property directly by solving the dynamical equation
(\ref{eq:3}) explicitly for the state-vector process
$\{|\psi_t\rangle\}$ in terms of independently specifiable random
data. We have also obtained closed form expressions for the energy
process $\{H_t\}$ and the reduction probability process
$\{\pi_{it}\}$. An argument based on a time-change shows that the
finite-time collapse model is related to the infinite-time
collapse model by a nonlinear transformation of the time variable.

\vskip1pc DCB thanks the Royal Society for support. We are
grateful to I.~Constantinou, T.~Dean, and J.~Dear for useful
discussions.

\end{document}